\documentclass[twocolumn]{revtex4}

\usepackage{graphicx,color}
\newcommand{\D}{\mathrm{d}}

\newcommand{\e}{\mathrm{e}}

\newcommand{\vecr}{(\mathbf{r})}

\newcommand{\kb}{k_{\mathrm{B}}}
\newcommand{\kbt}{k_{\mathrm{B}}T}

\newcommand{\lb}{l_\mathrm{B}}
\newcommand{\ld}{\lambda_\mathrm{D}}
\newcommand{\lgc}{\lambda_\mathrm{GC}}
\newcommand{\lgcm}{\lambda^\mathrm{m}_\mathrm{GC}}
\newcommand{\ls}{\lambda_\mathrm{s}}

\newcommand{\phia}{\phi_{\mathrm{A}}}
\newcommand{\phib}{\phi_{\mathrm{B}}}
\newcommand{\epsa}{\varepsilon_{\mathrm{A}}}
\newcommand{\epsb}{\varepsilon_{\mathrm{B}}}
\newcommand{\epsz}{\varepsilon_{0}}
\newcommand{\epsr}{\varepsilon_{\mathrm{r}}}
\newcommand{\epsavg}{\varepsilon_{\mathrm{av}}}

\newcommand{\alppm}{\alpha_{\pm}}

\begin{document}
%%%%%%%%%%%%%%%%

%%%%%%%%%%%%%%%%%%%%%%%%%%%%%%%%%%%%%%%%%%%%%%%%%%%%
\title{Ions in Mixed Dielectric Solvents:
Density Profiles and Osmotic Pressure between Charged Interfaces}
%%%%%%%%%%%%%%%%%%%%%%%%%%%%%%%%%%%%%%%%%%%%%%%%%%%%

\author{Dan Ben-Yaakov}
\author{David Andelman} \email{andelman@post.tau.ac.il}
\affiliation{Raymond and Beverly Sackler School of Physics and Astronomy, Tel Aviv
University, Ramat Aviv, Tel Aviv 69978, Israel}
\author{Daniel Harries}
\affiliation{Institute of Chemistry and The Fritz Haber Research Center, The Hebrew
University, Jerusalem 91904, Israel}
\author{Rudi Podgornik} \affiliation{Department of Theoretical Physics, J. Stefan
Institute, and Department of Physics, Faculty of Mathematics and Physics, University of Ljubljana, 1000 Ljubljana, Slovenia,}
\affiliation{Laboratory of Physical and Structural Biology, Eunice Kennedy Shriver National Institute of Child
Health and Human Development, National Institutes of Health, Bethesda, Maryland, 20814--0924}

\date{February 7, 2009}%
%after referees' comments
%final resubmitted version10

%%%%%%%%%%%%%%%%
\begin{abstract}

The forces between charged macromolecules, usually given in terms of osmotic pressure, are highly affected by the intervening ionic solution. While in most theoretical studies the solution is treated as a homogeneous structureless dielectric medium, recent experimental studies concluded that, for a bathing solution composed of two solvents (binary mixture), the osmotic pressure between charged macromolecules is affected by the binary solvent composition. By adding  local solvent composition terms to the free energy, we obtain a general expression for the osmotic pressure, in  planar geometry and within the mean-field framework. The added effect is due to the permeability inhomogeneity and nonelectrostatic short-range interactions between the ions and solvents (preferential solvation). This effect is mostly pronounced at small distances and leads to a reduction in the osmotic pressure for macromolecular separations of the order 1--2\,nm. Furthermore, it leads to a depletion of one of the two solvents from the charged macromolecules (modeled as planar interfaces). Lastly, by comparing the theoretical results with experimental ones, an explanation based on preferential solvation is offered for recent experiments on the osmotic pressure of DNA solutions.

\end{abstract}
%%%%%%%%%%%%%%%%

\maketitle

%\baselineskip=24pt

%%%%%%%%%%%%%%%%%%%%%%
\section{Introduction}
%%%%%%%%%%%%%%%%%%%%%%

The interactions between charged macromolecules immersed in aqueous solutions are of great
importance in biology and material science. Because of their relevance to colloidal
suspensions and biological macromolecules, the forces between charged objects mediated by
electrolytes have been the focus of numerous studies
\cite{forces1,forces2,forces3,Rau2000,Rau2004,Rau2006,preferential1,preferential2,textbooks}.
Within the context of the so-called primitive model, the solvent is modeled as a
homogeneous dielectric medium \cite{andelman2004, textbooks,henderson,netz_andelman}
affecting the system only through the dielectric constant that acts to reduce the
strength of the electrostatic field. More recently, theoretical and experimental
approaches, studying interactions between charged macromolecules, have been
extended to also treat binary solvent mixtures \cite{onuki1,onuki2,onuki3,tsori,abrashkin2007}.

The thermodynamics of binary solutions is well understood and has been described in
detail in many textbooks \cite{rubinstein_book}. However, there are still open questions
concerning the behavior of binary solutions in the presence of other degrees of freedom,
such as dissolved ions and external electric fields. These additional couplings are
relevant to a broader spectrum of applications, extending from manipulation of
microfluids \cite{microfluids1,microfluids2,microfluids3,microfluids4} to
biologically motivated problems such as protein stability and conformational changes \cite{biological_experiments1,biological_experiments2,biological_experiments3,biological_experiments4,daniel_review}.
For example, in recent experiments, the transition of a DNA
molecule from elongated coil to compact globule was found to depend on the addition of
another polarizable solvent to the aqueous solution \cite{Yoshikawa2005}, suggesting that
the interaction between DNA segments is modified by the presence of this additional solvent.

The effects of adding cosolvents to aqueous solutions of charged macromolecules can be
quite pronounced.  In fact, one of the common ways to precipitate DNA involves adding an
excess amount of ethanol to the aqueous solution, which counteracts
the repulsion between charged DNA strands \cite{Rau2000,Rau2004,Rau2006}.
This effect has been commonly attributed to the change in solution
dielectric constant. However, studies over the last decade convincingly
demonstrated that alcohol changes the disjoining (\textit{i.e.}, the
interaction) pressure between DNA strands to a much greater extent
than would be expected from the direct change in the dielectric constant.
This added effect that goes beyond changing of the dielectric constant
has been explained in terms of the preferential exclusion of alcohol
from the vicinity of interacting DNA strands \cite{Rau2006}. These
studies further demonstrated that because alcohol exclusion causes
an additional osmotic pressure difference
between the bulk solution and the concentrated DNA phase, DNA
strands are pushed even closer together.

Two distinct features prevail when trying to model ions immersed in a binary solvent mixture
within the standard Poisson--Boltzmann (PB) theory. First, the disparity between the solvent
permeabilities leads to a {\it dielectrophoretic} force. The ensuing force acts on the solution,
attracting the high permeability solvent (e.g., water) component toward the charged macromolecular
surface and, at the same time, depleting the lower permeability one (e.g., alcohol). As a result, the
solution becomes inhomogeneous and a permeability gradient is created in the vicinity of
the charged interface, where the system favors the higher permeability component that can
better screen the electrostatic field while excluding the low permeability solvent
away from charged interfaces. The second important feature is the chemical
(nonelectrostatic) preference of the ions for one of the two solvents. The
dissolved ions effectively drag with them a solvation shell preferentially
enriched in one of the solvents, thus repelling the second. When attracted to
the oppositely charged surfaces, the dissolved ions thereby change the composition
of the vicinal solvent. These two effects can enhance or compensate each other.
In this work, we treat only the case where the two effects act synergistically
to mutually enhance each other.

Previous theoretical works describing the effects of binary solvent mixtures dealt mainly
with systems close to their critical point. Tsori and Leibler investigated the change in the
phase transition temperature due to dielectric inhomogeneity and preferential solvation
\cite{tsori}, while Onuki and Kitamura investigated corresponding surface tension and the ionic distribution
near an interface \cite{onuki1,onuki2,onuki3}. To contrast and compare,
in the present work, we focus on binary solution
systems in the single phase region and away from the coexistence region. Moreover,
contrary to previous works, our main interest is the effect of the dielectrophoretic force
and preferential solvation on the pressure (or forces) between two equally charged
objects, such as a pair of charged DNA strands.

We model the system by delimiting ourselves to the simple planar geometry
for two interacting macromolecular surfaces. Some experimental setups apply
directly to this geometry and even for more complex setups, our model captures
the essential physics of coupling between the binary solvents and mobile ions.

In what follows, we present a model where the ionic densities, the solvent relative
composition, and the electrostatic potential are all continuous functions of the
local position. We derive a set of coupled differential equations relating the
various degrees of freedom at thermal equilibrium. Furthermore, we derive a general
expression for the local pressure in the form of a modified contact theorem 
and provide proof that it is spatially homogeneous.  This allows us to reduce
the corresponding Poisson--Boltzmann equation to a first order differential
equation that greatly simplifies the numerical problem.

Our numerical and analytical results focus on the solution mixtures where
the low permeability solvent (alcohol) has a small concentration compared to the other
solvent (water). First, we examine the influence of the various parameters on the
composition profiles.  We find that the deviation of the solvent composition profile from
its average (bulk) value can lead to large deviations from the regular Poisson--Boltzmann
theory predictions, especially regarding interactions between charged macromolecular
surfaces.  We investigate in detail the pressure dependence on the interplate separation
and its sensitivity to controllable parameters, such as salt concentration and average
solvent composition. Finally, we show a comparison between our pressure
profiles and the relevant experiment on DNA \cite{Rau2006}.

%%%%%%%%%%%%%%%%%%%
\section{The Model}
%%%%%%%%%%%%%%%%%%%

%Introduce the model with some general remarks.
%%%%%%%% Schematic figure %%%%%%%%%%%%%
\begin{figure}
\includegraphics[width=0.45\textwidth]{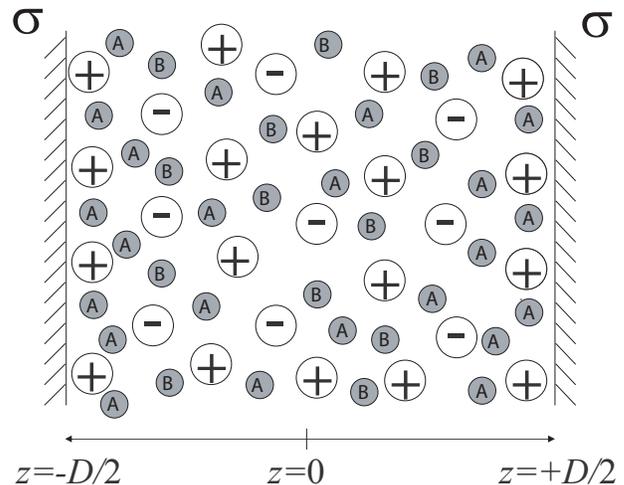}
\caption{Schematic illustration of the model system. The two
plates residing at $z=\pm D/2$ are charged with surface charge
density $\sigma<0$. The two solvents are represented by circles
denoted A and B, with dielectric constants $\varepsilon_\mathrm{A}$
and $\varepsilon_\mathrm{B}<\varepsilon_\mathrm{A}$, respectively.
The counterions and the high dielectric solvent ($\epsa$) are
attracted to the plates.} \label{fig1}
\end{figure}
%%%%%%%%%%%%%%%%%%%%%%%%%%%%

In the model considered here, ions are immersed in a binary mixture
of two solvents confined between two planar charged interfaces. The
two surfaces carry homogeneous surface charge densities, $\sigma<0$ (see Figure.~\ref{fig1}). The model is formulated on a mean-field level but modifies the regular PB theory in two important aspects.
First, the volume fractions of the two solvents, $\phia$ and
$\phib=1-\phia$, are allowed to vary spatially. Consequently, the
dielectric permeability of the binary mixture is also a function of
the spatial coordinates. In the following, we assume that the local
dielectric response $\varepsilon\vecr$ is a (linear) compositionally
weighted average of the two permeabilities $\epsa$ and $\epsb$:
\begin{equation}
\label{linear_eps0}
\varepsilon\vecr=\phia\vecr\epsa+\phib\vecr\epsb\, ,
\end{equation}
or
\begin{equation}
\label{linear_eps}
\varepsilon\vecr=\epsz-\phi\vecr\epsr\, ,
\end{equation}
where we define $\phi\equiv\phib$, $\epsz\equiv\epsa$ and $\epsr\equiv\epsa-\epsb$. This
linear interpolation assumption not only is commonly used but also is supported by experimental
evidence \cite{linear_dielectric1,linear_dielectric2}. Note that the incompressibility
condition satisfies $\phia+\phib=1\,$, meaning that the ionic volume fractions are neglected.

Apart from long-range electrostatic interactions between dissolved
ionic species, we also consider the case where short-range
interactions make an important contribution to equilibrium
properties. Consequently, pairwise short-range interactions between
all constituents contribute additional terms to the total free
energy and modify their equilibrium distributions, as will be
elaborated below.

The equilibrium properties are derived  within the mean-field framework.  Thermodynamic
equilibrium is obtained by minimizing the (grand canonical) thermodynamic potential,
$G=\int \D^3 r\, g\vecr$, leading to a generalized PB equation that also contains the
contribution of short-range interactions. The force equilibrium leading to interactions
between the confining surfaces is then obtained from the first integral of the PB equation
and can be reduced to a surface-normal term of the generalized stress tensor evaluated at
the bounding surfaces. The spatial profiles of the two solvents and ions, as well as the
equilibrium forces, can be obtained from a variational principle of the thermodynamic
potential.

%%%%%%%%%%%%%%%%%%%%%%%%
\subsection{Free Energy}
%%%%%%%%%%%%%%%%%%%%%%%%

We write the bulk free energy as a sum of four terms:
\begin{equation} \int_V\D^{3}r\left[f_{\rm e}\vecr+f_{\rm i}\vecr+f_{\rm m}\vecr+f_{\rm s}\vecr\right]\,,
\label{freeen-1}
\end{equation}
where the free energy density $f=f_{\rm e}+f_{\rm i}+f_{\rm m}+f_{\rm s}$ and $V$ is the total volume.
The first term is due to electrostatic interactions between
the ionic species mediated by the dielectric medium and characterized by the spatially
inhomogeneous dielectric function $\varepsilon\vecr$. For simplicity, the dissolved ions are assumed
to result from a completely dissociated (1:1) monovalent salt. In this case, the
electrostatic term, $f_{\rm e}$, is given by
\begin{equation}\label{FE_e}
{f_{\rm e}}=-\frac{\varepsilon\vecr}{8\pi}({\bf \nabla}\psi)\,^{2}+e(
n_{+}-n_{-}) \psi\, ,
\end{equation}
where $\psi\vecr$ is the electrostatic potential, $e$ is the electron charge, and
$n_{\pm}\vecr$ are the number densities (per unit volume) of the monovalent co- and counterions. Note that
the first term implicitly couples the electric field, $\mathbf{E}=-{\bf \nabla}\psi$, with the
solvent composition, $\phi\vecr$, via the spatial dependence of the dielectric
response, $\varepsilon\vecr$, on local composition as was defined in eq~\ref{linear_eps}. This
is the dielectrophoretic term mentioned previously that favors a higher
local dielectric constant (lower $\phi\vecr$), and causes
attraction of water to the charged surface. This is one of
the two sources of the composition inhomogeneity in the model.

The entropy of ion mixing constitutes the second term, $f_{\rm i}$, given by
\begin{equation}\label{FE_i}
f_{\rm i}= \kbt \bigg[ n_{+}\left(\log (n_{+}a^3)-1
\right)+n_{-}\left(\log (n_{-}a^3)-1\right)\bigg]\, .
\end{equation}
where $\kb$ is the Boltzmann constant, $T$ is the temperature, and $a^3$ is the molecular
volume. The third term, $f_{\rm m}$, accounts for the binary-mixture free energy given in our model by regular
solution theory:
\begin{equation}\label{FE_bm}
f_{\rm m}=\frac{\kbt}{a^3}\left[\phi\log \phi+(1-\phi)\log (1-\phi)+\chi\phi(1-\phi)\right] \, .
\end{equation}
The first two terms represent the solvent entropy of mixing, while the third represents
the bilinear short-range interactions between the two solvents. The interaction parameter,
$\chi$, is dimensionless (rescaled by $\kbt$). Note that we took the same molecular
volume $a^3$ for both A and B components. In general, this is not a serious deficiency and
can be easily amended, if necessary.

The fourth term, $f_{\rm s}$, in the free energy originates from the preferential interaction of the
ions with one of the two solvents as described in the Introduction. We assume here that
this preference can be described by a bilinear coupling between the two ion densities
$n_\pm$ and the relative solvent composition $\phi$, which is the lowest
order term that accounts for these interactions (higher order correlation terms can be
added in a systematic way). The preferential solvation energy, $f_{\rm s}$, is then given by
\begin{equation}\label{FE_s}
f_{\rm s}=\kbt \left(\alpha_{+} n_{+}+\alpha_{-} n_{-}\right)\phi\, ,
\end{equation}
where the dimensionless parameters $\alpha_\pm$ describe the solvation preference of the ions,
defined as the difference between the solute (free) energies dissolved in the A and B
solvents. The free energy $f_{\rm s}$ corresponds closely to the Gibbs energy of transfer from one solvent to another,
as is described in detail in refs~\cite{marcus1,marcus2}.
This bilinear coupling represents a second source of composition inhomogeneity.
Namely, a density profile of the ions $n_\pm\vecr$ (a diffusive layer near a charged
object) forces a corresponding solvent profile, $\phi\vecr$.

To all these bulk terms, one must add a surface term, describing the electrostatic
interactions between charged solutes and confining charged interfaces. This surface term
is given by
\begin{equation}
F_A= \oint_{A} \D^{2}r \, e \sigma \psi_{\rm s},
\end{equation}
where $\psi_{\rm s}$ is the electrostatic potential evaluated at the bounding surfaces and depends on the surface charge density $\sigma$ (charged groups per unit area) and
surface area $A$. Note that the charged surface is described by a uniform
charge density, $\sigma$. In a more refined model, nonelectrostatic interactions,
such as preferential adsorption
of the two solvents, and finite ion
effects could be included as well.

The total free energy is then written as a sum of the bulk and surface terms
\begin{equation}
\int_{V}\D^{3}r f (\psi,  n_{\pm},\phi) + \oint_{A} \D^{2}r\, e \sigma \psi_{\rm s}\,.
\label{totalfreeenergy}
\end{equation}
In the grand-canonical ensemble the corresponding thermodynamic potential is given by
\begin{equation}
g\vecr=f\vecr-\kbt\left[\mu_{+}
n_{+}\vecr+\mu_{-} n_{-}\vecr+\mu_{\phi}\frac{\phi\vecr}{a^3}\right] \,,
\label{totalpot}
\end{equation}
where $\mu_{\pm}$ and $\mu_{\phi}$ are the dimensionless chemical potentials coupled to
the ionic densities $n_\pm$ and the relative solvent composition $\phi$, respectively.

In thermodynamic equilibrium, the spatial profile of the various degrees of freedom
characterizing the system is obtained by deriving the appropriate Euler--Lagrange (EL)
equations via a variation principle of the thermodynamic potential, eq~\ref{totalpot}.
The EL equations are then reduced to four coupled differential equations for the four degrees
of freedom, $\psi\vecr$, $n_{\pm}\vecr$ and $\phi\vecr$:
\begin{equation}
\label{3d_eom_psi} {\bf \nabla}\cdot\left(
\frac{\varepsilon}{4\pi}{\bf \nabla}\psi\right)+e(n_{+}-n_{-})=0
\end{equation}
\begin{equation}
\label{3d_eom_n} \pm \frac{e\psi}{\kbt}+\log(n_{\pm}a^3)+\alpha_{\pm}\phi-\mu_\pm=0
\end{equation}
\begin{eqnarray}
\label{3d_eom_phi} \log
\left(\frac{\phi}{1-\phi}\right)+\frac{\epsr a^3}{8\pi\kbt}({\bf
\nabla}\psi)^2
 +\chi(1-2\phi) \nonumber\\ +~ a^3\left(\alpha_{+} n_{+} + \alpha_{-} n_{-}\right)
-\mu_\phi=0\, .
\end{eqnarray}
At the charged interfaces, an additional equation stems from the surface term of
eq~\ref{totalfreeenergy}
\begin{equation}
\label{BC}
\frac{\delta g}{\delta\psi_{s}}=0 \hspace{0.3cm} \Rightarrow \hspace{0.3cm} \hat{{\bf n}}
\cdot {\bf \nabla}\psi\bigg{|}_s = -\frac{4\pi e}{\varepsilon_{\rm s}}\sigma,
\end{equation}
where $\hat{{\bf n}}$ is the unit vector normal to the bounding surfaces, $\varepsilon_{\rm s}=\epsz-\epsr \phi_{\rm s}$ and $\phi_{\rm s}$ are the surface values of $\varepsilon$ and $\phi$, respectively. The last
equation, just as for standard PB theory, expresses the electroneutrality
of the system, as can be shown by the integral form of Gauss law.

By solving the above set of equations, one can obtain the spatial profiles of the various
degrees of freedom at thermodynamic equilibrium. For a general geometry, these equations
can be solved only numerically.

%%%%%%%%%%%%%%%%%%%%%%%%%%%%
\subsection{Bulk Behavior}
%%%%%%%%%%%%%%%%%%%%%%%%%%%%

In the bulk, the system is homogeneous having a zero potential $\psi=0$
and bulk values of $n_{+}=n_{-}=n_{\rm b}$
and $\phi=\phi_{\rm b}$. The EL equations, eqs~\ref{3d_eom_psi}-\ref{3d_eom_phi}, reduce to:
\begin{eqnarray}
\log(n_{\rm b} a^3) + \alpha_{\pm}\phi_{\rm b}-\mu_\pm=0\hspace{3.4cm}\nonumber\\
\log
\left(\frac{\phi_{\rm b}}{1-\phi_{\rm b}}\right)+\chi(1-2\phi_{\rm b})+
a^3(\alpha_{+} + \alpha_{-}) n_{\rm b}
-\mu_\phi=0\, .\nonumber\\
\end{eqnarray}
Eliminating the $n_\pm$ fields, we remain with a single bulk equilibrium
equation
\begin{equation}
\label{phi_bulk} \log
\left(\frac{\phi_{\rm b}}{1-\phi_{\rm b}}\right)+\chi(1-2\phi_{\rm b})+ \Gamma \e^{-
\frac{1}{2} (\alpha_{+} + \alpha_{-}) \phi_{\rm b} } -\mu_\phi=0\, ,
\end{equation}
where $\Gamma$ is defined as
\begin{equation}
\Gamma = (\alpha_{+} + \alpha_{-}) \e^{\frac{1}{2}(\mu_{+} + \mu_{-})}.
\end{equation}
Depending on the values of $\mu_\phi\,$, $\mu_\pm\,$, $\alpha_\pm\,$, and $\chi$, the
solutions of the bulk equation correspond either to a single phase of density $\phi_{\rm b}$ or to a
coexistence between two phases with different densities. Hereafter,
we restrict ourselves to the single-phase region of the phase diagram,
where the chemical potentials $\mu_\pm$, $\mu_{\phi}$ follow directly
from the form of the bulk free energy.

%%%%%%%%%%%%%%%%%%%%%%%%%%%%
\subsection{Planar Geometry}
%%%%%%%%%%%%%%%%%%%%%%%%%%%%

% Deriving the pressure. And the 1D equations (optional).

We exploit the symmetry of a planar system in order to derive analytically the pressure acting on the boundaries of the confined system. For a binary mixture confined to a slab delimited by two planar charged surfaces of infinite lateral extent (see Figure \ref{fig1}), the general treatment introduced above can be simplified, and the free energy can be cast into a one-dimensional integral over the normal $\hat{z}$ direction. For this special case, we show next that the pressure is proportional to the first integral of the EL equations. Using this expression, we also derive a first-order differential equation for the electrostatic potential that will greatly simplify the problem.

%%%%%%%%%%%%%%%%%%%%%%%%%%%%%%%%%%%%%%%%%%%%%%%
\subsubsection{Pressure in Planar Geometries}
%%%%%%%%%%%%%%%%%%%%%%%%%%%%%%%%%%%%%%%%%%%%%%%

We start from a general form of the free energy $F$ which depends on $N$ one-dimensional
fields $\left\{\psi_1(z),...\, ,\psi_N(z)\right\}$ and their derivatives $\left\{\psi'_1(z),...\,
,\psi'_N(z)\right\}$
\begin{equation}
F/A=\int \D z f\left( \left\{
\psi_i(z),\psi'_i(z)\right\};z\right)\, .
\end{equation}
When $f$ does not depend explicitly on the coordinate $z$, ${\partial f}/{\partial z}=0$,
we obtain the following relation (see Appendix):
\begin{equation}\label{eq_g}
f-\sum_{i=1}^N\frac{\partial f}{\partial
\psi'_i}\psi'_i\,=\,\mathrm{const}\, .
\end{equation}
In our problem, $f$ can be written as a sum of electrostatic and nonelectrostatic contributions
\begin{equation}\label{grand_potetnial}
f=-\frac{\varepsilon\left( \{n_i\}\right)}{8\pi}\psi'^2+\sum_{i=1}^{N}
q_i n_i \psi + h \left( \{ n_i \} \right)\, ,
\end{equation}
where $h$ is the grand potential of $N$ different species with densities $\{ n_1,\,
...\, ,n_N\}$ of a general form but without any electrostatic
interactions. The charge of the $i$th species is
denoted by $q_i$, and $\varepsilon\left(\{n_i\}\right)$ is the
dielectric response as a function of the densities $\{ n_1,\,...\, ,n_N\}$.
Substituting eq~\ref{grand_potetnial} into eq~\ref{eq_g}, we obtain
\begin{equation}
\label{const_motion}
-\frac{\varepsilon }{8\pi}\psi'^2+\sum_{i=1}^{N} q_i n_i
\psi + h  +\frac{\varepsilon }{4\pi}\psi'^2=\mathrm{const}\, .
\end{equation}
Finally, using the equilibrium equations for the densities $\{ n_1, ...\, ,n_N\}$
\begin{equation}
\frac{\partial f }{\partial n_i}=-\frac{1}{8\pi}\frac{\partial
\varepsilon }{\partial
n_i}\psi'^2+q_i\psi+\frac{\partial h  }{\partial n_i}=0\, ,
\end{equation}
we end up with the following expression
\begin{equation}
\frac{1}{8\pi}\left[\varepsilon +\sum_i\frac{\partial \varepsilon }
{\partial n_i}n_i\right]\psi'^2+h  -\sum_i n_i\frac{\partial h  }{\partial n_i}=\mathrm{const}\, .
\label{firstintegral}
\end{equation}
For the special case of non-charged liquid mixtures, $f$ reduces
to $h$, while it follows from
general thermodynamic identities \cite{rowlinson} that the last
two terms in eq~\ref{firstintegral} are
equal to the negative of the local pressure
\begin{equation}
P= -h  +\sum_i n_i\frac{\partial h }{\partial n_i} \,.
\end{equation}
However, even in a charged liquid mixture the electrostatic
potential vanishes away from the boundaries so that $P$ is also the
bulk value of the pressure in a charged system. Together with
eq~\ref{const_motion}, it follows that the first integral can be
cast into the form:
\begin{equation}
 - P=\frac{1}{8\pi}\left[\varepsilon +\sum_i\frac{\partial \varepsilon }
 {\partial n_i}n_i\right]\psi'^2+h  -\sum_i n_i\frac{\partial h  }{\partial n_i},
\label{firstintegral-1}
\end{equation}
Namely, the integration constant of eq~\ref{firstintegral} is simply
the negative of the pressure, and is a constant throughout the
system. We next consider separately the properties of the
electrostatic and non-electrostatic terms in eq~\ref{firstintegral}.

The first term is nothing but the negative of the $zz$ component of the Maxwell
electrostatic stress tensor, appropriately generalized to the case where the dielectric
permeability is density dependent~\cite{LL}. The last two terms together, as already
noticed, represent the local pressure of the system in the presence of charges. In the
standard PB theory, these two terms are given by the van't Hoff form, while here they are
given by an appropriate generalization, stemming from the free energy {\sl ansatz},
eq~\ref{freeen-1}. Combining all the terms in eq~\ref{firstintegral}, we get the total
$zz$ component of the stress tensor, which in thermodynamic equilibrium has to be a
constant and equal to {-}$P$, eq \ref{firstintegral-1}.

Note that the above proof is valid for any form of the free energy
$f$ (as $h$ had an arbitrary form), and accounts for electrostatic
as well as non-electrostatic degrees of freedom in a completely
general way. Applying this general result to our free energy,
eqs~\ref{freeen-1}-\ref{FE_s}, yields the following form of the
total pressure:
\begin{eqnarray}
\label{pi_total} P
=-\frac{1}{8\pi}\left(\varepsilon_0-2\varepsilon_r
\phi\right)\psi'^2\hspace{3cm}\nonumber
\\ +\kbt\left(n_+ + n_- -
\frac{\log(1-\phi)}{a^3}\right) \nonumber \hspace{0.2cm}\\ +
\kbt\left(\alpha_+ n_+ \phi +\alpha_-  n_- \phi
-\frac{\chi\phi^2}{a^3}\right)\,.\hspace{-0.3cm}
\end{eqnarray}
This pressure $P$ should be compared with the pressure of the standard PB theory:
\begin{eqnarray}
P_\mathrm{PB} =-\frac{1}{8\pi} \varepsilon\psi'^2+\kbt\left(n_+ + n_-\right)\, ,
\end{eqnarray}
and contains several additional terms. In fact, the difference between the two is
twofold: first, a polarizability term of the form
$\varepsilon_r\phi\psi'^2/4\pi$ is included in the
pressure since the dielectric constant is now spatially dependent.
This term is equal to the product of the polarization $-\epsr
\phi\psi'$ and the electric field $E =-\psi'$. Second, the
short-range interactions also change the form of the pressure:
the solvent interactions contribute the term $-\chi\phi^2/a^3$, and the
ion--solvent interactions add the two terms, $\alpha_+ n_+ \phi$ and $\alpha_- n_- \phi$.
This last addition changes the pressure significantly when considering two similarly
charged surfaces. We will discuss this point at length below.

%%%%%%%%%%%%%%%%%%%%%%%%%%%%%%%%%%%%%%%%%%%%%%%%%%%%
\subsubsection{First Integral of the EL Equations in Planar Geometries}
%%%%%%%%%%%%%%%%%%%%%%%%%%%%%%%%%%%%%%%%%%%%%%%%%%%%

We now use the form of the first integral of the EL equations (eq~\ref{pi_total}) to
obtain an explicit first-order differential equation for the electric field. The EL
equations for the ion densities, eq~\ref{3d_eom_n}, give the following relations:
\begin{equation}
n_\pm(\psi,\phi)=n_{\rm b} \e^{\mp e \psi/\kbt-\alpha_\pm(\phi-\phi_{\rm b})}\,.
\end{equation}
  From the first integral, we now deduce
\begin{eqnarray}
\label{ode_1d} \left(\frac{\D \psi}{\D z}\right)^2 =
\frac{8\pi\kbt}{(\varepsilon_0-2\varepsilon_r\phi)}\bigg({n}_+
+{n}_- -2n_{\rm b}\nonumber \hspace{2cm}\\ \left.+\alpha_+({n}_+ \phi
-\phi_{\rm b} n_{\rm b})+\alpha_-({n}_- \phi -
\phi_{\rm b} n_{\rm b})\right.\nonumber\hspace{0.6cm}\\
\hspace{0.5cm}+\frac{1}{a^3}\log{\frac{1-\phi_{\rm b}}{1-\phi}}
\left.-\frac{1}{a^3}\chi(\phi^2-\phi_{\rm b}^2)-n_{\rm b}\Pi \right)\,
.\hspace{0.2cm}
\end{eqnarray}
%
%where $x=z/\lambda_0$ is the dimensionless coordinate scaled by $\lambda_0=1/\sqrt{8\pi
%e^2 n_{\rm b}/\kbt}$.Note that $\lambda_0$ is \textit{not} the regular Debye length,
%$\lambda_\mathrm{D}$, as it has no dielectric constant dependence.
The difference between the pressure $P$ at finite separation and its bulk value
$P_b$ (infinite separation) is given by the rescaled osmotic pressure $\Pi=(P-P_b)/\kbt n_{\rm b}$ and
\begin{equation}
\frac{P_b}{\kbt n_{\rm b}}=2+(\alpha_+ +\alpha_-)\phi_{\rm b}
-\frac{1}{a^3n_{\rm b}}\log(1-\phi_{\rm b})-\frac{1}{a^3n_{\rm b}}\chi\phi_{\rm b}^2\,.
\end{equation}
For a single plate (or, equivalently in the limit of two plates at infinite separation),
$\Pi$ vanishes. Note that ${n}_\pm$ in eq~\ref{ode_1d} are functions of
$\phi$ and $\psi$, and $\phi$ is by itself a function of
$\psi$ and $\psi'$, given by eq~\ref{3d_eom_phi} that is a
transcendental algebraic equation for $\phi$:
\begin{eqnarray}
\label{1d_eom_phi} \frac{\varepsilon_r}{8\pi\kbt} \left(\frac{\D
\psi}{\D z}\right)^2+\alpha_+( {n}_+-n_{\rm b})+\alpha_-( {n}_-
- n_{\rm b})\nonumber \hspace{0.5cm}\\
+\frac{1}{a^3}\left(\log{\frac{\phi}{1-\phi}}-
\log{\frac{\phi_{\rm b}}{1-\phi_{\rm b}}} -2\chi(\phi-\phi_{\rm b})\right)=0\, .
\end{eqnarray}

The boundary conditions for each plate{\slash}boundary are given by three coupled
algebraic equations for $\phi_{\rm s}$, $\psi_{\rm s}$ and $\psi_{\rm s}'$. The first two equations are
eqs~\ref{ode_1d} and \ref{1d_eom_phi}. The third equation is given by the
electroneutrality condition, eq~\ref{BC}, that can be simply rewritten in the form
\begin{equation}
\label{BC_dimensionless}
\varepsilon_{\rm s}\psi_{s}'\,+\,4\pi e \sigma =0\, ,
\end{equation}
and $\varepsilon_{\rm s}$ was defined after eq~\ref{BC}.
%%%%%%%%%%%%%%%%%%%%%%%%%%%%%%%%
\section{Results and Discussion}
%%%%%%%%%%%%%%%%%%%%%%%%%%%%%%%%

The equilibrium equations eqs~\ref{ode_1d}, \ref{1d_eom_phi}, and \ref{BC_dimensionless}
derived above have no closed analytical solution. Hence, we solve them numerically
to obtain spatial profiles for $\phi$ and $n_\pm$. In addition, by considering the
new terms as small perturbations (compared to the regular PB theory),
we show that an approximate analytical solution can be derived for the
single plate case in the absence of salt. We then show numerical
results for the pressure as a function of separation and for the
pressure dependence on the experimentally controlled parameters
$\alpha_+$, $\phi_{\rm b}$, and $n_{\rm b}$. Lastly, we compare our results to
one available set of experiments
on DNA in a binary solvent mixture.

%%%%%%%%%%%%%%%%%%%%%%%%%%%%%
\subsection{Density and Permeability Profiles}
%%%%%%%%%%%%%%%%%%%%%%%%%%%%%

% The effect of inhomogeneous dielectric response and $\chi$
% parameter on the profiles is negligible. See Figure~\ref{fig2}.

We investigate the limit of small concentrations of the low
dielectric component with no preferential solvation interactions
($\alpha_\pm=0$) and for two values of the $\chi$ parameter. In
Figure~\ref{fig2}, we compare these numerical solutions for a single
surface (for which the osmotic pressure vanishes, $\Pi=0$, similar
to infinite interplate separation) to the ones of the regular PB
equation with a homogeneous (average) dielectric constant, $\epsavg
\equiv\epsa -\phi_{\rm b}(\epsa-\epsb)$.

When all additional interactions are omitted ($\alppm=0$ and
$\chi=0$), the difference between the two models is negligible.
While the deviation in $\phi$ right at the surface reaches $10\%$ of
its bulk value $\phi_{\rm b}$, it leads to only a $~0.5\%$ deviation for
the dielectric constant $\varepsilon$ at the surface. The dependence
of the other fields $\psi$ and $n_\pm$ on $\phi$ is only due to
changes in the dielectric constant. Therefore, in the limit of no
short-range interactions, these fields hardly differ from the
results of the regular PB model with homogeneous dielectric constant
$\epsavg$. The correction due to addition of solvent short-range
interactions is also found to be small, even for larger
solvent--solvent interaction, $\chi=1.5$. This $\chi$ value still
describes a single bulk phase, as it is smaller than the critical
value $\chi_\mathrm{c}=2$. We conclude that, in the absence of
preferential solvation, $\alpha_\pm=0$, the modified PB model has
only a small added effect on the permeability, as can be seen in
Figure~\ref{fig2}.

%%%%%%%% Fig 2 %%%%%%%%%%%%%
\begin{widetext}\phantom{}
\begin{figure}[t!]
\centerline{ \includegraphics[width=0.9\textwidth]{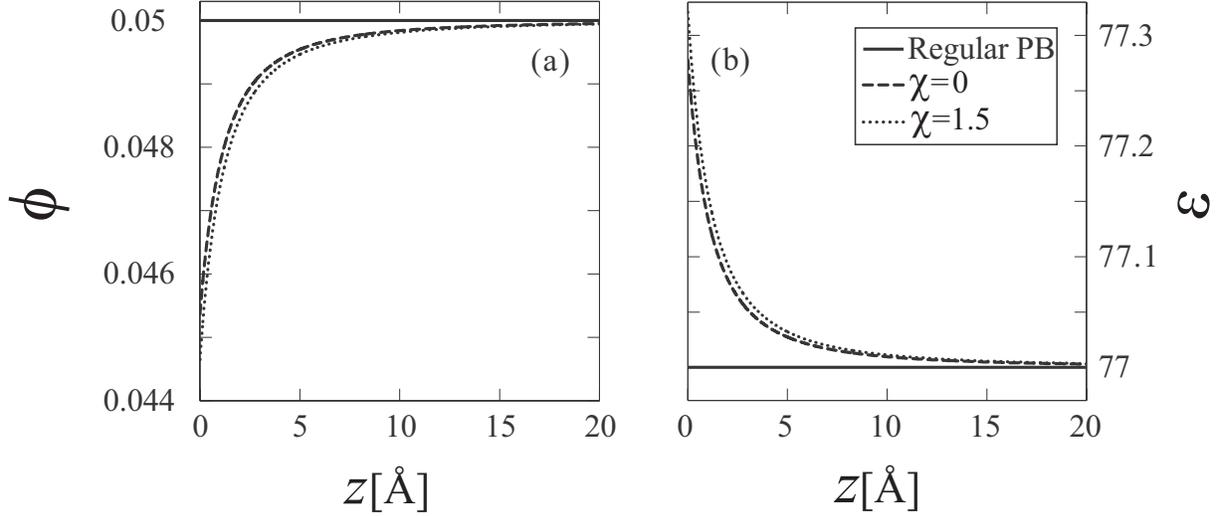} }
\caption{Spatial profiles of (a) the solvent relative composition
$\phi$ and (b) the permeability $\varepsilon$. The regular PB with
homogeneous dielectric constant $\varepsilon=77$ (solid line) is
compared with our modified PB for binary mixture with and without
short-range interactions, $\chi=0$ (dashed line) and $\chi=1.5$
(dotted line), respectively. Other parameters are:
$\sigma=-1/100$\AA$^{-2}$, $n_{\rm b}=10^{-4}$M, $\varepsilon_\mathrm{A}=80$,
$\varepsilon_\mathrm{B}=20$ and $\phi_{\rm b}=0.05$. In all the cases, no
preferential solvation is considered, $\alpha_\pm=0$.} \label{fig2}
\end{figure}
\end{widetext}
%%%%%%%%%%%%%%%%%%%%%%%%%%%%

% The effect of $\Delta\alpha_\pm$ on the profiles. See
% Figure~\ref{fig3}.

In Figure~\ref{fig3}, we examine numerically the effect of preferential solvation on the
solvent profile in the low concentration limit ($\phi_{\rm b}=0.09$).
Simply stated, when the ions prefer to be in the vicinity of the high permeability solvent
molecules, we expect an increase in the exclusion of the low permeability solvent near
the wall. Indeed, as can be seen in Figure~\ref{fig3}, the exclusion of the low
permeability solvent increases with $\alpha_+$. We also find that the value
of the co-ion--solvent short-range interaction parameter $\alpha_-$ does not significantly
affect the form of the permeability profile and is set hereafter to zero.

Moreover, when increasing $\alpha_+$ even further ($\alpha_{+}=30$ in Figure~\ref{fig3}), $\phi(z)$ and $\varepsilon(z)$ have a sharp change at about $z=2$\AA. Namely, a layer rich in A species is formed near the wall with
a thickness of a few angstroms, and at a certain distance from the wall $\phi$ decreases abruptly to a value close to
the bulk value $\phi_{\rm b}$. This phenomenon is clearly a consequence of the $n_+\phi$ coupling and is in itself not an electrostatic effect. A gradient squared term $\left(\nabla\phi\right)^2$ in the free energy would have smoothed out this behavior and will be considered elsewhere. The steep variation observed for the nonhomogeneous mixture could be due to the fact that the boundary condition demands a $\phi$ value that is much different from the one that follows from the bulk equation (eq~\ref{phi_bulk}), so that the system tends to exhibit bulklike behavior as soon as possible. A detailed analysis of this phenomena will be presented in a separate study.

%%%%%%%% Fig 3 %%%%%%%%%%%%%
\begin{widetext}\phantom{}
\begin{figure}[t!]
\centerline{ \includegraphics[width=0.93\textwidth]{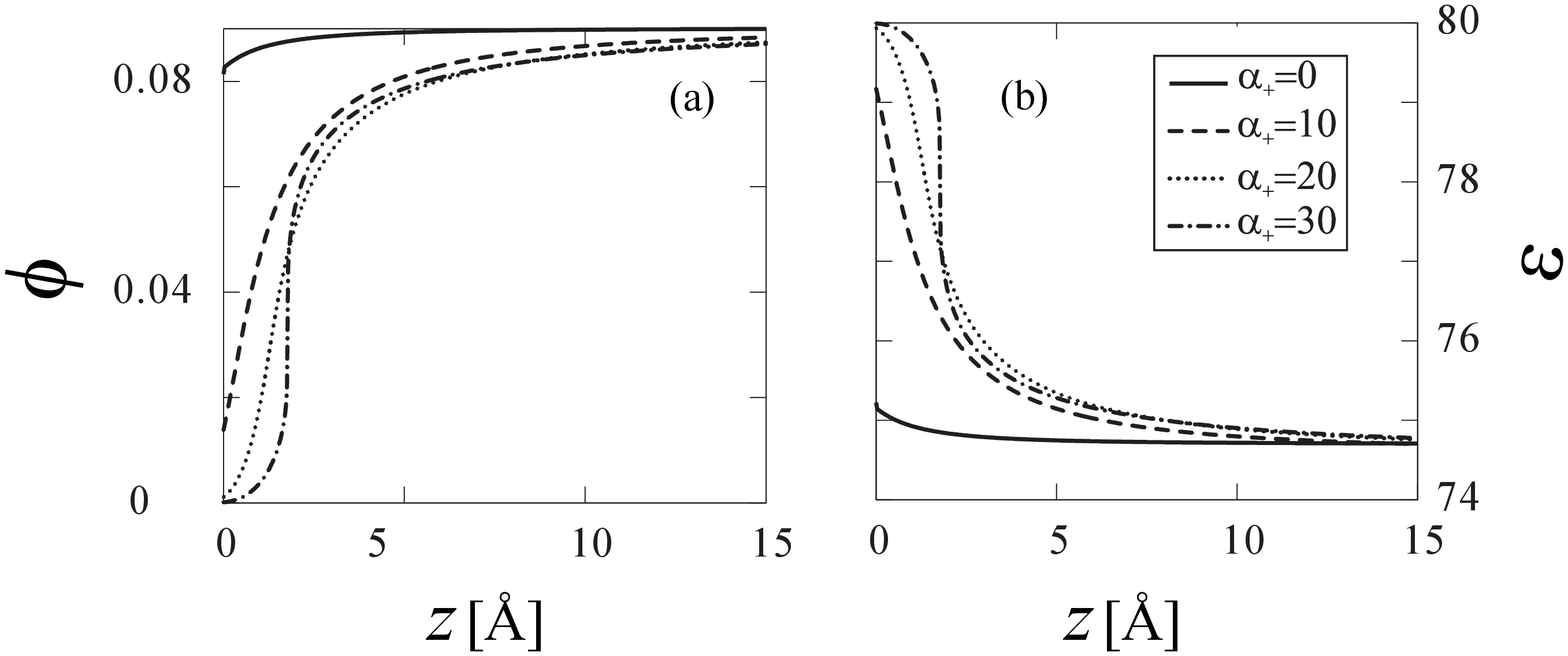} }
\caption{Spatial profiles of (a) the solvent composition $\phi$ and
(b) the dielectric constant $\varepsilon$ for various values of
$\alpha_+$ as shown in the legend. The co-ion parameter is $\alpha_-=0$, the surface charge
density is $\sigma=-1/100$\AA$^{-2}$, and the salt concentration is
$n_{\rm b}=10^{-4}$M. The solvent dielectric constants are
$\varepsilon_\mathrm{A}=80$ and $\varepsilon_\mathrm{B}=20$. The
bulk value of $\phi$ is $\phi_{\rm b}=0.09\,$.} \label{fig3}
\end{figure}
\end{widetext}
%%%%%%%%%%%%%%%%%%%%%%%%%%%%

%%%%%%%%%%%%%%%%%%%%%%%%%%%%%%%%%%%%%%%%%%%%%%%%%%%%%%%%%%%%%%%%%%%%
\subsection{Some Analytic Results}
%%%%%%%%%%%%%%%%%%%%%%%%%%%%%%%%%%%%%%%%%%%%%%%%%%%%%%%%%%%%%%%%%%%%

The model described above can be solved analytically by making some
simplifying assumptions, and considering certain limiting behaviors.
We assume that there are no solvent short-range interactions
($\chi=0$) and that the preferential solvation interaction is weak
compared to $\kbt$ ($\alpha_+\ll1$). We also assume that the
contribution due to the permeability inhomogeneities  is negligible,
and take $\varepsilon$ throughout the system to be the weighted
average of the two bulk relative compositions, $\epsavg$. Moreover,
we take the limit of zero salt, as if only counterions are present
to keep the system neutral. Lastly, we assume that one of the
solvent concentrations is much smaller than the other.

With these assumptions, we can practically isolate the effect of
preferential solvation and obtain analytical profiles. The PB
equation in this limit assumes the form
\begin{equation} \frac{\D^2\psi}{\D
z^2}=-\frac{4\pi e n(z)}{\varepsilon_{\rm av}}\, ,
\end{equation}
where the ion density $n(z)$ is a function of both
the potential $\psi$ and the solvent relative composition $\phi$
\begin{equation}
 \label{no_salt_n} {n(z)=n_{\rm b}\e^{-e \psi/\kbt-\alpha(\phi-\phi_{\rm b})}} .
\end{equation}
The prefactor $n_{\rm b}$ is determined by satisfying the electroneutrality
condition, and the subscript {\footnotesize $\pm$} in $n_\pm$ is omitted in this counterion only case.

The composition $\phi$ as a function of $n$ is
\begin{equation}
\phi=\phi_{\rm b}\e^{-\alpha a^3 n}\simeq\phi_{\rm b}(1-a^3\alpha n)\, .
\end{equation}
Here, we made use of the assumption that the preferential solvation interaction is small,
$\alpha(a^3 n)\ll1$. Substituting it back into eq~\ref{no_salt_n},
we obtain for the further limit,\linebreak $\alpha \phi\ll1\,$,
\begin{equation}
n=\frac{n_{\rm b} \e^{-e \psi/\kbt}}{1-\alpha^2a^3\phi_{\rm b}
n_{\rm b}\e^{-e \psi/\kbt}}.
\end{equation}
For finite values of $e\psi/\kbt$ and in the limit of $\alpha^2\phi_{\rm b}
a^3 n_{\rm b} \ll 1$, we obtain to lowest order
\begin{equation}
n\simeq
n_{\rm b}\e^{-e \psi/\kbt}\left(1+\alpha^2\phi_{\rm b}
a^3n_{\rm b}\e^{-e \psi/\kbt}\right).
\end{equation}
Using this equation in the PB equation, we get a second order
differential equation for $\psi$:
\begin{equation}
\frac{\D^2\psi}{\D z^2}=-\frac{4\pi e n_{\rm b}}{\varepsilon}\left(\e^{-e \psi/\kbt}+\alpha^2\phi_{\rm b}
a^3n_{\rm b}\e^{-2e \psi/\kbt}\right)\, .
\end{equation}
For a single plate, the boundary condition to the equation above is given by
\begin{equation}
\frac{e}{\kbt}\frac{\D \psi}{\D z}\bigg|_{z=0}=4\pi \lb |\sigma|\equiv\frac{2}{\lgc}\, ,
\end{equation}
where $\lb=e^2/\varepsilon_{\rm av}\kbt$, and $\lgc$ is the well known Gouy--Chapman (GC) length \cite{andelman2004}.

Solving the equation above, we obtain the following results:
\begin{eqnarray}
\psi(z)=\frac{\kbt}{e}\log\left[(z+\lgcm)^2-\lambda_s^2\right]+\psi_0 \,,\\
{n(z)=\frac{(z+\lgcm)^2+\ls^2}{2\pi\lb\left[(z+\lgcm)^2-\ls^2\right]^2}}\, ,\hspace{1.7cm}
\end{eqnarray}
where $\ls^2=\alpha^2 a^3 \phi_{\rm b}/4\pi\lb$ is a typical
length associated with the $\alpha$ parameter. The second length scale is the modified Gouy--Chapman
(GC) length $\lgcm$ obtained by satisfying the boundary
condition:
\begin{equation}
\lgcm=\frac{\lgc}{2}\left(1+\sqrt{1+4(\ls/\lgc)^2}\,\right)\,.
\end{equation}
For $\ls\ll\lgc$ one obtains $\lgcm\simeq\lgc\left[1+(\ls/\lgc)^2\right]$. Thus,
the effect of preferential solvation enhances the ion density in the proximity of the surface and results in a faster decay of the density profile.

%%%%%%%%%%%%%%%%%%%%%%%%%%%%%%%%%%%%%%%%%%
\subsection{Pressure vs. Separation Curves}
%%%%%%%%%%%%%%%%%%%%%%%%%%%%%%%%%%%%%%%%%%

Due to symmetry, the pressure between two identically charged plates
is most conveniently calculated at the midplane. When considering
the change in pressure due to the permeability inhomogeneity, one
can separate the direct and indirect corrections. The direct one is
due to the change of midplane composition $\phi(z)$, while the
indirect one is related to changes of the midplane ion density. In
the absence of preferential solvation ($\alpha_\pm=0$), $\phi$ at
the midplane depends only on the local value of the electrostatic
field (see eq~\ref{1d_eom_phi} with $\alpha_\pm=0$). However, in a
symmetric twoplate system, the electrostatic field vanishes at the
midplane and $\phi$ there equals to its bulk value,
ultimately contributing no correction to the pressure. Moreover,
since $\varepsilon(z)$ turns out to be nearly homogeneous (see
Figure~\ref{fig2}), the indirect correction is minute as well. Thus,
in the absence of preferential solvation the combined effect of a
binary mixture is negligible.

When adding the preferential solvation term characterized by the
parameter $\alpha_\pm$ (the term coupling between $n_\pm$ and
$\phi$), $\phi$ becomes dependent on the nonzero midplane
potential. As a consequence, the midplane $\phi$ value differs from
$\phi_{\rm b}$, and results in two direct corrections to the pressure. The
first comes from the osmotic pressure of the solvent
($\sim\kbt(\phi-\phi_{\rm b})/a^3)$, while the second originates from the
$\phi\cdot n_\pm$ coupling term. Thus, even if the indirect
contribution to the profiles is negligible, the direct correction
alone can substantially change the pressure.

%%%%%%%% Fig 4 %%%%%%%%%%%%%
\begin{figure}
\includegraphics[width=0.39\textwidth]{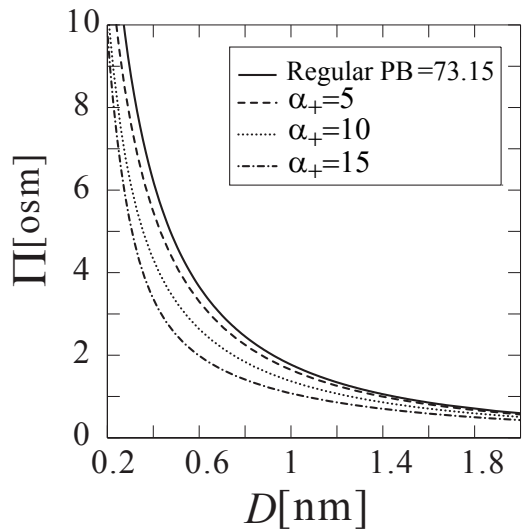}
\caption{\label{fig4}Dependence of pressure on separation $D$
for various ion--solvent interaction strengths $\alpha_+$ as shown in the legend. Other
parameters are $\sigma=-1/100$\AA$^{-2}\,$, $n_{\rm b}=10^{-4}$M,\\
 $\varepsilon_\mathrm{A}=80$, $\varepsilon_\mathrm{B}=4$, and
$\phi_{\rm b}=0.09\,$.}
\end{figure}
%%%%%%%%%%%%%%%%%%%%%%%%%%%%

The effect of preferential solvation on the pressure between two
identically charged surfaces is now examined for various values of
the $\alpha_+$ parameter. As the coupling term in eq~\ref{pi_total}
contributes directly to the pressure, the effect of changing
$\alpha_+$ is rather pronounced in the pressure versus separation
curves (see Figure~\ref{fig4}). The figure clearly shows that the
pressure decreases when $\alpha_+$ increases. Yet, because the
effect is short ranged, the major differences are observed at small
separations ($D<2$nm). From the analysis of the profiles with
respect to the interaction strength (Figure~\ref{fig4}), we conclude
that the change in pressure is substantial only up to distances of a
few nanometers. This means that for large separations the midplane
values of the fields $\psi$, $n$ and $\phi$ within our model will be
similar to the regular PB theory predictions. However, for small
separations of the order of 1--2 nanometers, both the profiles and
the pressure are affected by the preferential solvation.

%%%%%%%%%%%%%%%%%%%%%%%%%%%%%%%%%%%%%%%%%%%%%%%%%%%%%%%%%
\subsection{The Effect of the Solution Parameters on the Pressure}
%%%%%%%%%%%%%%%%%%%%%%%%%%%%%%%%%%%%%%%%%%%%%%%%%%%%%%%%%

%%%%%%%% Fig 5 %%%%%%%%%%%%%
\begin{figure}
\includegraphics[width=0.39\textwidth]{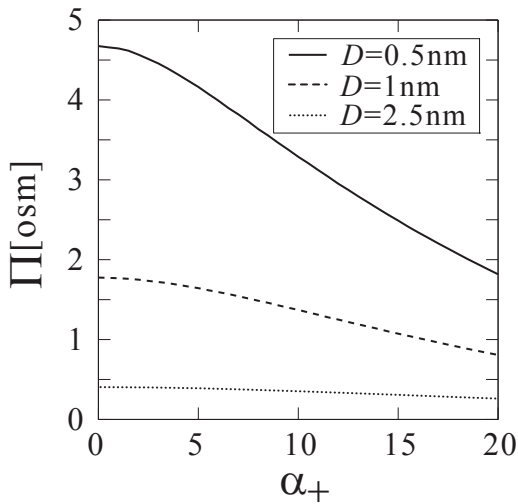}
\caption{\label{fig5}Pressure as a function of $\alpha_+$ for various interplate
separations: $D=0.5,1,$ and $2.5\,$nm. Other parameters are
 $\sigma=-1/100$\AA$^{-2}\,$, $n_{\rm b}=10^{-5}$M, $\varepsilon_\mathrm{A}=80$, $\varepsilon_\mathrm{B}=4$, and \\$\phi_{\rm b}=0.09\,$.}
\end{figure}
%%%%%%%%%%%%%%%%%%%%%%%%%%%%

We proceed by examining the influence of the parameters $\alpha_+$, $\phi_{\rm b}$ and $n_{\rm b}$
on the pressure. The $\alpha_+$ parameter can be modified experimentally by using
different solvents, while $\phi_{\rm b}$ and $n_{\rm b}$ can be easily controlled
in the experiment by changing composition.

In Figure~\ref{fig5}, we present the dependence of the pressure $\Pi(\alpha_+; D)$
on the interaction strength $\alpha_+$ for a fixed separation. As expected, for
small values of $\alpha_+$ ($<2$), the pressure depends only weakly on the interaction
strength, whereas for larger values the pressure falls with $\alpha_+$. This
implies that there is a value of $\alpha_+$ where its direct contribution
to the pressure becomes larger than all other contributions (electrostatic
and entropic). Moreover, the slope of $\Pi(\alpha_+; D)$ depends on the
separation, as can be clearly seen by comparing the $D=0.5\,$nm and $D=1\,$nm
results. For smaller separations ($D=0.5\,$nm), the preferential solvation
effect is stronger in accordance with the results shown in the previous
sections (Figures~\ref{fig2}--\ref{fig4}), where the effect has a range
of a few nanometers. For $D=2.5\,$nm, the pressure changes very slowly
with $\alpha_+$ and the preferential solvation is small even for large $\alpha_+$ ($15<\alpha_+<20$).

%%%%%%%% Fig 6 %%%%%%%%%%%%%
\begin{figure}
\includegraphics[width=0.39\textwidth]{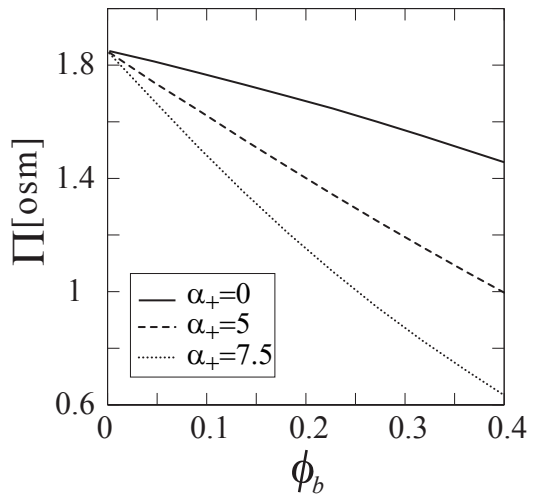}
\caption{\label{fig6}Pressure versus bulk solvent composition $\phi_{\rm b}$
for various ion--solvent interaction strength, $\alpha_+$. Other parameters
are $\sigma=-1/100$\AA$^{-2}\,$, $n_{\rm b}=10^{-5}$M, $\varepsilon_\mathrm{A}=80\,$,
and $\varepsilon_\mathrm{B}=4\,$. The separation is fixed at $D=1$nm.}
\end{figure}
%%%%%%%%%%%%%%%%%%%%%%%%%%%%

Next, we investigate how increasing $\phi_{\rm b}$ changes
the pressure. In Figure~\ref{fig6}, we plot the pressure versus $\phi_{\rm b}$
for a fixed separation $D=1\,$nm for three values of the interaction
strength $\alpha_+$. Increasing the low permeability solvent concentration
decreases the pressure through decrease of the permeability and by
increasing the preferential solvation. The results in Figure~\ref{fig6}
suggest that even for $\alpha_+=0$ (no preferential solvation)
the pressure decreases with $\phi_{\rm b}$, implying that the dielectrophoretic
mechanism contributes a nearly linear dependence on $\varepsilon_r$.
When increasing $\alpha_+$, the  $\Pi(\phi_{\rm b})$
slope is steeper due to higher bulk pressure of the low
permeability solvent, contributing directly to the pressure
through the preferential solvation term (eq~\ref{pi_total}).

Since there is a linear $\phi_{\rm b}$ term in the pressure \linebreak($\sim-\kbt \phi_{\rm b}/a^3$),
one can deduce from the results above that the main contribution to the pressure
comes simply from a higher reference concentration $\phi_{\rm b}$ that reduces
the pressure. Namely, for small separations, the solvent $\phi$ and the ion $n_\pm$
at the midplane have only a weak dependence on $\phi_{\rm b}$, similar to regular PB
theory where the pressure has no dependence on the bulk
salt density $n_{\rm b}$ at small separations \cite{andelman2004}.

%%%%%%%% Fig 7 %%%%%%%%%%%%%
\begin{figure}
\includegraphics[width=0.39\textwidth]{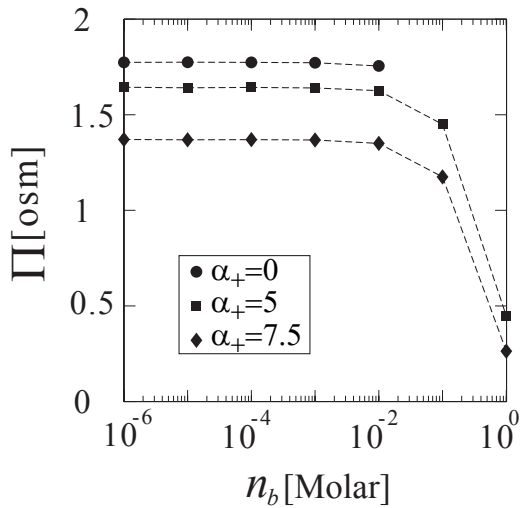}
\caption{\label{fig7}Pressure versus bulk salt concentration $n_{\rm b}$ for
various ion--solvent interaction strength, $\alpha_\pm$. Other parameters
are $\sigma=-1/100$\AA$^{-2}\,$, $\varepsilon_\mathrm{A}=80$, $\varepsilon_\mathrm{B}=4$,
and $\phi_{\rm b}=0.09\,$. The separation is fixed at $D=1$nm. Lines are guides for the eye.}
\end{figure}
%%%%%%%%%%%%%%%%%%%%%%%%%%%%

Finally, we investigate the influence of the salt concentration on the pressure at fixed
separation ($D=1$nm). The results are presented in Figure~7, where we plot the pressure
versus the salt concentration for fixed separation and for different values of the
interaction strength $\alpha_+$. For low concentration ($n_{\rm b}<10^{-2}\,$M), the pressure
has no dependence on $n_{\rm b}$. It is known from the regular PB theory \cite{andelman2004}
that, at small separations, when the Debye length is much larger than the separation ($\ld\gg
D$), the pressure only weakly depends on the salt concentration. In this sense, the
modified PB theory presented here is similar to the regular PB theory. The effect
of preferential solvation is just a constant addition to the pressure, which keeps the
same dependence of the pressure on the salt concentration. When going to larger concentrations,
where the Debye length becomes comparable to or smaller than the separation, the pressure
decays exponentially as can be expected also from standard PB theory.

%%%%%%%%%%%%%%%%%%%%%%%%%%%%%%%%%%%%%%%%%%
\subsection{Comparison of Model with Experiment}
%%%%%%%%%%%%%%%%%%%%%%%%%%%%%%%%%%%%%%%%%%

So far, we have shown that we are able to account for some of the
forces that lead to cosolvent preferential exclusion from charged
interacting macromolecular surfaces. In particular, our model
accounts not only for locally varying dielectric profiles that
follow the solvent mixture composition (through the variables
$\varepsilon_{\rm A}$ and $\varepsilon_{\rm B}$), but also for preferential ion
solvation (through the variables $\alpha_\pm$), that in turn depends
on local solvent composition as well. Using these two sets of
parameters, it is possible to propose a physical mechanism for
solute (or solvent) exclusion from interacting surfaces. The
different dielectric constants of the two solvent components cause
depletion of one of the components from the charged surface. This
variation in solvent composition can in turn affect the
concentration of ions between the two interacting surfaces. The
combined effect can lower the disjoining pressure between equally
charged surfaces by varying local solution concentrations.

While it is hard to unambiguously prove the origins of the molecular interactions that lead to the different solvation properties of ions and cosolvent in the vicinity of complex macromolecules, we show that by varying the model parameters we can account for the observed trends in the experimental studies of Rau and Stanley \cite{Rau2006}. In these experiments, osmotic pressure is applied to a condensed phase of DNA strands in aqueous solution by adding a
neutral polymer, poly(ethylene glycol) (PEG) that is completely excluded from the DNA phase. The DNA--DNA spacings in solution, $D$, are measured using small angle X-ray scattering. In addition, salt and different alcohols are added at different concentrations. Figure~\ref{fig8}a shows the experimentally derived $\Pi(D)$ (equation of state) for DNA solutions containing either 0.02 or 1.2\,M NaBr salts, as they appear in ref~\cite{Rau2006}. Solutions to which 2-methyl-2,4-pentanediol (MPD) alcohol was added at concentrations of 0.5 or 1\,M are compared with solution with no MPD. As the figure shows, for both salt concentrations, the added salt lowers the DNA--DNA spacing for a particular applied osmotic pressure. However, the figure also clearly shows that the reduction in $\Pi$ is more significant at the higher salt concentration. This would imply that salts (or more generally electrostatic forces) are involved in
determining the effect of MPD on $\Pi$, suggesting an important role reserved for the dielectric properties that can be linked to the distribution of ions and cosolvent partitioning in between the DNA strands.

%%%%%%%% Fig 8 %%%%%%%%%%%%%
\begin{widetext}\phantom{}
\begin{figure}[t!]
\centerline{ \includegraphics[width=0.9\textwidth]{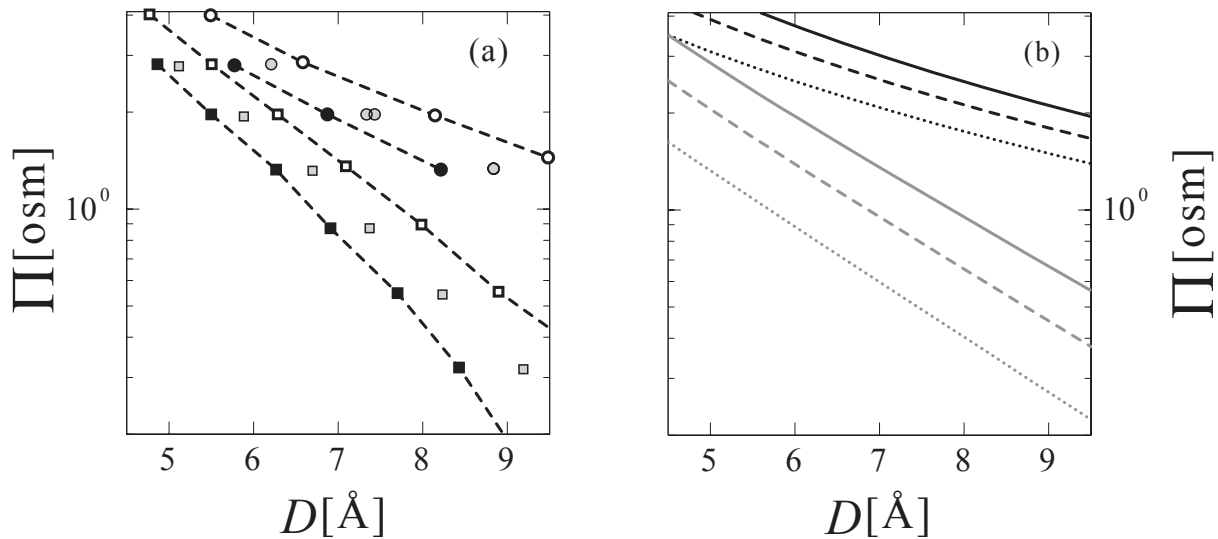}}
\caption{Pressure $\Pi$ as a function of the separation $D$. (a)
Experimental data. Circles (squares) represent the 0.02\,M
(1.2\,M) added NaBr salt results adopted from ref~\cite{Rau2006}.
Empty circles and squares represent the experiment with no MPD. Gray
circles and squares represent the experiment with 0.5\,M MPD. Black
circles and squares represent the experiment with 1\,M MPD. The dashed
lines are guides to the eye. (b) Numerically calculated lines from
the model with parameters taken to match the experiment. The
ion--solvent interaction strength is the same for all lines
$\alpha_+=5\,$, and it was treated as a fitting parameter. The surface
charge is taken as $\sigma=-1/100$\AA$^{-2}\,$ to fit the DNA values.
The dielectric constants are $\varepsilon_\mathrm{A}=80$ (water) and
$\varepsilon_\mathrm{B}=25$ (MPD). Black lines are for
$n_{\rm b}=0.02$\,M and gray lines are for $n_{\rm b}=1.2$\,M. Solid, dashed, and
dotted lines stand for $\phi_{\rm b}=0.0018\,$, $\phi_{\rm b}=0.126\,$, and
$\phi_{\rm b}=0.252\,$, respectively.} \label{fig8}
\end{figure}
\end{widetext}
%%%%%%%%%%%%%%%%%%%%%%%%%%%%

Even though our model considers interactions between flat surfaces, rather than the cylindrical ones expected for DNA strands, we show that it is possible to get similar trends to those found in experiment using the two sets of the $\epsilon$ and $\alpha$ parameters. Figure~\ref{fig8}b shows our results for two charged plates with charge density similar to that of DNA ($\sigma=-1/100$\AA$^{-2}$), and dielectric constants of $\epsilon_A=80$ (representing water) and $\epsilon_B=25$ (close to the value of pure MPD). In our model, we use only the relative volume fraction $\phi_{\rm b}$ of the two species, water and MPD, without accounting for their different molar weight (water molar weight is 18 and that of MPD is 118). In the experiment \cite{Rau2006}, the two MPD solutions have concentrations of 0.5 and 1.0\,M corresponding, respectively, to $\phi_{\rm b}=0.055$ and $0.11$. These values are in good agreement with the values chosen in our model (Figure~\ref{fig8}b) to give a good fit to the experimental data: $\phi_{\rm b}=0.126$ and 0.252. Note that the value $\phi_{\rm b}=0.0018$ is chosen for convenience to fit the zero MPD concentration case. The value of $\alpha_+$ could, in principle, be evaluated from the solvation free energy of NaBr salts in binary water--MPD solutions of different contents. Because such data is lacking, we treat $\alpha_+$ as a fitting parameter and use $\alpha_+=5$ (in units of $\kbt$), allowing us to fit closely the experimental data. We note that this $\alpha_{+}$ value is close to the experimental transfer free energy of sodium from water to ethanol $\sim 5\kbt$ as was quoted in ref~\cite{marcus2}. The value of $\alpha_{-}$ for bromide, which was taken in our work as zero, seems to be generally lower than that of $\alpha_{+}$ but is less clearly resolved~\cite{marcus1}.

We find that for this set of parameters the numerical calculation {\sl grosso modo}
follows the experimental trends: the spacing between curves grows with equal additions of
alcohol to the solution, but for the higher salt concentration the change in $\Pi$ is larger.
These results underscore two important general conclusions. First, it would be impossible to explain
the difference in the $\Pi(D)$ behavior for high salt versus low salt without discussing ions and
the role of the dielectric medium. Our model introduces these species in a self-consistent
manner through the PB-like theory. Second, the comparison demonstrates the important role
reserved for the preferential ion-solvation interactions in the different components of
the binary solution. Specifically, it would be impossible to explain the shifts in
distances at a given applied osmotic stress without invoking a nonzero $\alpha_+$, that in
turn shifts the exclusion of ions due to changes in solution composition in the DNA phase.

%%%%%%%%%%%%%%%%%%%%%
\section{Conclusions}
%%%%%%%%%%%%%%%%%%%%%

The model presented in this work is a modification of the regular PB theory and accounts for the effects of a dielectric medium composed of two solvents. There are two important features that modify the standard PB theory. Due to different dielectric constants of the two solvents, the permeability is no longer homogeneous. This effect is accounted for by coupling the electrostatic field to the solvent local composition. The second modification is the addition of a preferential solvation term which enables the ions to drag a favorable solvent and locally deplete the other. This is modeled by a coupling term between the ion density and the solvent local composition. Similar models can be used for a number of problems such as the properties of interfaces \cite{onuki2,onuki3}, critical behavior of binary mixture in the presence of ions \cite{tsori,onuki1}, and the forces between more elaborated charged macromolecules, which is the main focus of this work.

For two identically charged planes, we find that, in the absence of preferential solvation, the density profiles and the pressure undergo only small modifications. This is demonstrated numerically and supported by an argument that relies on the symmetry of the two-plate system. However, by adding the preferential solvation term, we are able to observe a considerable correction to the pressure at small separations. The coupling between ion density and solvent local composition appreciably changes the midplane concentration values, and as a consequence the pressure is reduced. We also investigated the dependence of the pressure on experimentally controlled parameters such as salt concentration, bulk solvent composition, and preferential solvation strength. The pressure depends on the preferential solvation
but changes substantially only at small separations (1--2~nm). The threshold of preferential solvation energy which is required to change the pressure significantly is on the order of a few $\kbt$. It is found that the pressure depends nearly linearly on the bulk relative composition, implying that the density profiles at small separations have no dependence on $\phi_{\rm b}$. Finally, the effect of added salt to the solution seems to change the pressure in the same manner as for regular PB theory.

We also used our model to put fourth an explanation for the experimentally
measured pressures in a condensed phase of DNA. The comparison shows that the experimental
trend is bourne out by our model results. This suggests that the main mechanism
causing the depletion of one solvent away from the charged macromolecule is very plausibly the
preferential solvation of the ions. Thus, beyond simple electrostatic screening, salt ions may
play an additional and important role in the behavior of charged macromolecules immersed in solution.

Further applications and refinements of the model could be considered. For example, the
model can be used to analyze the effect of strong preferential solvation on the critical
behavior. As shown in Figure~\ref{fig3}, the solvent relative composition profile becomes
discontinuous at strong preferential solvation. We believe that this can be
explained in the framework of a Ginzburg--Landau theory {that would} account for the phase
transition. Moreover, in addition to the simple planar geometry treated here, other geometries
such as a cylindrical one can obtain a more direct quantitative comparison with DNA experiments.
Finally, the limit of ionic dilute solutions can be generalized to the concentrated limit,
including the full entropy of mixing, as was considered in ref~\cite{henridavid}.

%%%%%%%%%%%%%%%%%%%%%%%%%%%%%%%%%%%%%%%%%%
\begin{acknowledgments}
%%%%%%%%%%%%%%%%%%%%%%%%%%%%%%%%%%%%%%%%%%

We thank Ariel Abrashkin,  Adrian Parsegian, and Yoav Tsori for numerous discussions. We are indebted to Don Rau, and Brian Todd for helpful discussions and for sharing experimental data with us. D.A. acknowledges support from the Israel Science Foundation (ISF) under Grant No. 231/08 and the U.S.-Israel Binational Foundation (BSF) under Grant No. 2006/055. R.P. and D.H. would like to acknowledge the support from the Israeli and Slovenian Ministries of Science through a joint Slovenian-Israeli research grant. The Fritz Haber research center is supported by the Minerva foundation, Munich, Germany. This study was in part supported by the Intramural Research Program of  the  NIH,  Eunice  Kennedy Shriver National  Institute  of  Child  Health  and  Human  Development.

%%%%%%%%%%%%%%%%%%%%%
\end{acknowledgments}
%%%%%%%%%%%%%%%%%%%%%

%%%%%%%%%% Captions %%%%%%%%%%%%%%%%%%%%%%%%%
%\newpage
%%%%%%%%%%%%%%%%%%%%%%%%%%%%%%%%%%%%%%%%%%%%%%%%%%%%%%%%%%%%%%%%%%%%
\appendix \section{Derivation of Pressure in One Dimensional System}
%%%%%%%%%%%%%%%%%%%%%%%%%%%%%%%%%%%%%%%%%%%%%%%%%%%%%%%%%%%%%%%%%%%%

In the following we present the derivation of eq.~\ref{eq_g}. We
start from a free energy $F$, eq~18 which depends on $N$ coupled one-dimensional fields
$\left\{\psi_1(z),...\, ,\psi_N(z)\right\}$ and their derivatives
$\left\{\psi'_1(z),...\, ,\psi'_N(z)\right\}$:
\begin{equation}
F/A=\int  f\left( \left\{
\psi_i(z),\psi'_i(z)\right\};z\right)\D z\, ,
\end{equation}
where $i=1,...\, ,N$. There are $N$ EL equations
\begin{equation}
\frac{\delta F}{\delta \psi_i}=0\hspace{0.3cm}\Rightarrow \hspace{0.3cm} \frac{\partial f }{\partial
\psi_i}-\frac{\D}{\D z}\frac{\partial f}{\partial \psi_i'}=0\, .
\end{equation}
The total derivative of $f(z)$ is
\begin{equation}
\frac{\D f }{\D z}=\sum_i\left(\frac{\partial f}{\partial
\psi_i}\psi'_i+\frac{\partial f}{\partial
\psi'_i}\psi''_i\right)+\frac{\partial f}{\partial z}\, . \label{eq_a}
\end{equation}
Moreover, we can write
\begin{equation}
\frac{\partial f}{\partial \psi'_i}\psi''_i=\frac{\D }{\D
z}\left(\frac{\partial f}{\partial \psi'_i}\psi'_i\right)-\frac{\D
}{\D z}\left(\frac{\partial f}{\partial \psi'_i}\right)\psi_i'\, .
\end{equation}
Substituting this back into eq \ref{eq_a} and using the EL equations
we find
\begin{equation}
\frac{\D f }{\D z}=\sum_i\left[\frac{\D }{\D z}\left(\frac{\partial
f}{\partial \psi'_i}\psi'_i\right)\right]+\frac{\partial f}{\partial
z}\, .
\end{equation}
When $f(z)$ does not depend explicitly on the coordinate $z$, $\frac{\partial f}{\partial
z}=0$, the last term vanishes and we end up with a first order differential relation:
\begin{equation}\label{eq_g_appendix}
f-\sum_i\left(\frac{\partial f}{\partial
\psi'_i}\psi'_i\right)=\mathrm{const}\, .
\end{equation}
%

%\newpage

%%%%%%%%%%%%%%%%%%%%%%%%%%%

%%%%%%%%%%%%%%%%%%%%%

%%%%%%%%%%%%%%
\end{document}